\documentclass[12pt,preprint]{aastex}
\def\chisq{$\chi^2$}

\def\kms{$\,$km$\,$s$^{-1}$}
\def\sqig{$\sim$}
\def\sun{$_\odot$}
\def\cts{counts~s$^{-1}$}
\def\degrees{$^{\circ}$}
\def\ergss{ergs~s$^{-1}$}
\def\ergscms{ergs~cm$^{-2}$~s$^{-1}$}

\def\src{XTE\,J1859+083}

\def\RXTE{{\it RXTE}}
\def\BeppoSAX{{\it BeppoSAX}}

\begin{document}

\title{\RXTE\ and \BeppoSAX\ Observations of the
Transient X-ray Pulsar XTE J1859+083}


\author{R.~H.~D.~Corbet\altaffilmark{1,2},
J.~J.~M.~in~'t~Zand\altaffilmark{3},
A.~M~Levine\altaffilmark{4},
F.~E.~Marshall\altaffilmark{5}}

\altaffiltext{1}{University of Maryland, Baltimore
County; corbet@umbc.edu}

\altaffiltext{2}
{CRESST/Mail Code 662,
NASA Goddard Space Flight Center, Greenbelt, MD 20771}

\altaffiltext{3}
{SRON Netherlands Institute for Space Research, Sorbonnelaan 2, 3584 CA
Utrecht, The Netherlands jeanz@sron.nl; Astronomical Institute,
Utrecht University, PO Box 80000, 3508 TA Utrecht, The Netherlands}

\altaffiltext{4}
{Kavli Institute for Astrophysics and Space Research, MIT,
Cambridge, MA 02139}

\altaffiltext{5}
{Mail Code 660.1,
NASA Goddard Space Flight Center, Greenbelt, MD 20771}

\begin{abstract}

We present observations of
the 9.8 s X-ray pulsar \src\ made with the ASM 
and PCA on board RXTE, and the WFC on
board \BeppoSAX. The ASM data cover a 12
year time interval and show that an extended outburst occurred between
approximately MJD 50,250 and 50,460
(1996 June 16 to 1997 January 12). The ASM data excluding this
outburst interval suggest a possible modulation with a period
of 60.65 $\pm$ 0.08 days. 
Eighteen sets of PCA observations were obtained over
an approximately one month interval in 1999. The flux variability
measured with the PCA appears consistent with the possible
period found with the ASM. 
The PCA measurements of the pulse period showed it to
decrease non-monotonically and then to increase significantly.
Doppler shifts due to orbital motion rather than accretion
torques appear to be better able to explain the pulse
period changes. 
Observations with the WFC during the extended outburst give a
position
which is consistent with a previously
determined PCA error box, but which has a significantly
smaller error. The transient nature of XTE J1859+083 and the length of its
pulse period are consistent with it being a Be/neutron star
binary. The possible 60.65 day orbital period would be of the expected
length for a Be star system with a 9.8 s pulse period.

\end{abstract}
\keywords{stars: individual (XTE J1859+083) 
--- stars: neutron ---
X-rays: binaries}

\section{Introduction}

The X-ray source \src\ was discovered by Marshall et al.
(1999) in observations made with the {\it Rossi X-ray Timing Explorer}
(\RXTE) Proportional Counter Array (PCA) on 1999
August 8 (MJD 51,398) during slews between pointed
observations of other targets. Although it was not possible
to search for pulsations in the slew observation,
Marshall
et al. (1999) found pulsations at a period of 9.801 $\pm$ 0.002 s
in an observation made on 1999 August 16.
Cross-scan observations with the PCA reported by
Marshall et al. (1999) located the source at
R.A. = 18$^h$59.1$^m$, 
decl. = +8\degrees\ 15' (equinox 2000.0), 
with an estimated uncertainty
of 2' (90\% confidence).
The transient nature of \src\ and its pulsations suggest
that it might be a member of the Be/neutron star
binary class of objects (e.g., Charles \& Coe 2006). 
Such sources are expected to vary on several timescales.
Periodic modulation of the X-ray flux on the orbital
period may occur if
the orbit of the neutron is eccentric. The system
may exhibit extended periods of quiescence with
no detectable X-ray flux when the circumstellar
envelope around the Be star disappears. Less frequently,
periods of exceptional brightness, known as Type II outbursts,
may also occur, possibly caused by an exceptional expansion
of the Be star envelope.

Since the discovery of this source, the only other
observations of \src\ that have been reported are by Romano et
al. (2007) who observed the PCA position on 
2007 November 16 to 17 (MJD 54,420.63 - 54,421.98) with the {\it Swift} X-ray Telescope.
Romano et al. (2007) did not detect the
source and reported an upper limit of
5$\times$10$^{-14}$ \ergscms\ in the energy range of 0.3 to 10 keV.
The non-detection of the source is consistent with
\src\ being a Be star source during its quiescent phase.

We present here the results of additional observations
with the \RXTE\ PCA, an improved position from \BeppoSAX\ Wide Field
Camera (WFC) observations, and an analysis of the \RXTE\ All Sky Monitor
(ASM)
light curve which reveals a possible orbital period.
The results of these observations are all consistent with
a Be/neutron star classification.

\section{Observations}

\subsection{\RXTE\ PCA}

The PCA is described in detail by Jahoda et al. (1996, 2006). This instrument
consists of five nearly identical Proportional Counter Units (PCUs)
sensitive to X-rays with energies between 2 and 60 keV with a total
effective area of 6500 cm$^2$. The Crab produces 13,000 \cts\ for the
entire PCA across the complete energy band. The PCA spectral
resolution at 6 keV is approximately 18\%, and the field of view is 
1\degrees\ FWHM.

After the first serendipitous slew observation,
pointed observations were carried out with two sets
of observations being made every few days over a period
of about 38 days.
The log of PCA observations is given in Table \ref{table:spectrum}.
We analyzed the \RXTE\ PCA observations using
standard procedures for background subtraction and light
curve and spectrum extraction.  In Table \ref{table:spectrum} 
we give 
the fluxes resulting from fitting the spectra obtained from
each observation
with the typical X-ray pulsar model of an absorbed power-law with
a high-energy cutoff (White et al. 1983). This model gave a good fit for all
observations and we see no large change in spectral parameters
as the flux declines with only possibly a small amount of steepening
of the spectral slope at lower fluxes.
Although some of the spectral parameters, particularly the 
absorption, are not well constrained at low flux levels
we adopt the same spectral model at all flux levels
for consistency.
The mean spectral parameters are found to
be: photon index = 0.84 $\pm$ 0.03, $N_{\rm H}$ = 2.1 $\pm$ 0.2
$\times$ 10$^{22}$ cm$^{-2}$, $E_{\rm cut}$ = 6.41 $\pm$ 0.06 keV,
and $E_{\rm fold}$ = 13.0 $\pm$ 0.3 keV.

We extracted lightcurves from standard 1 data, which have 0.125 sec
time resolution, and no energy selection. We made barycenter
timing corrections using {\em faxbary}  and combined
observations
that were close to each other 
to obtain improved period measurements if phase connection
was possible. We used an epoch folding technique to 
determine pulse period from the resulting 7 light
curves. Errors were obtained 
by determining the periods at which 
the \chisq\ value was 1 less than the peak
value.

\begin{figure}
\epsscale{0.9}
\plotone{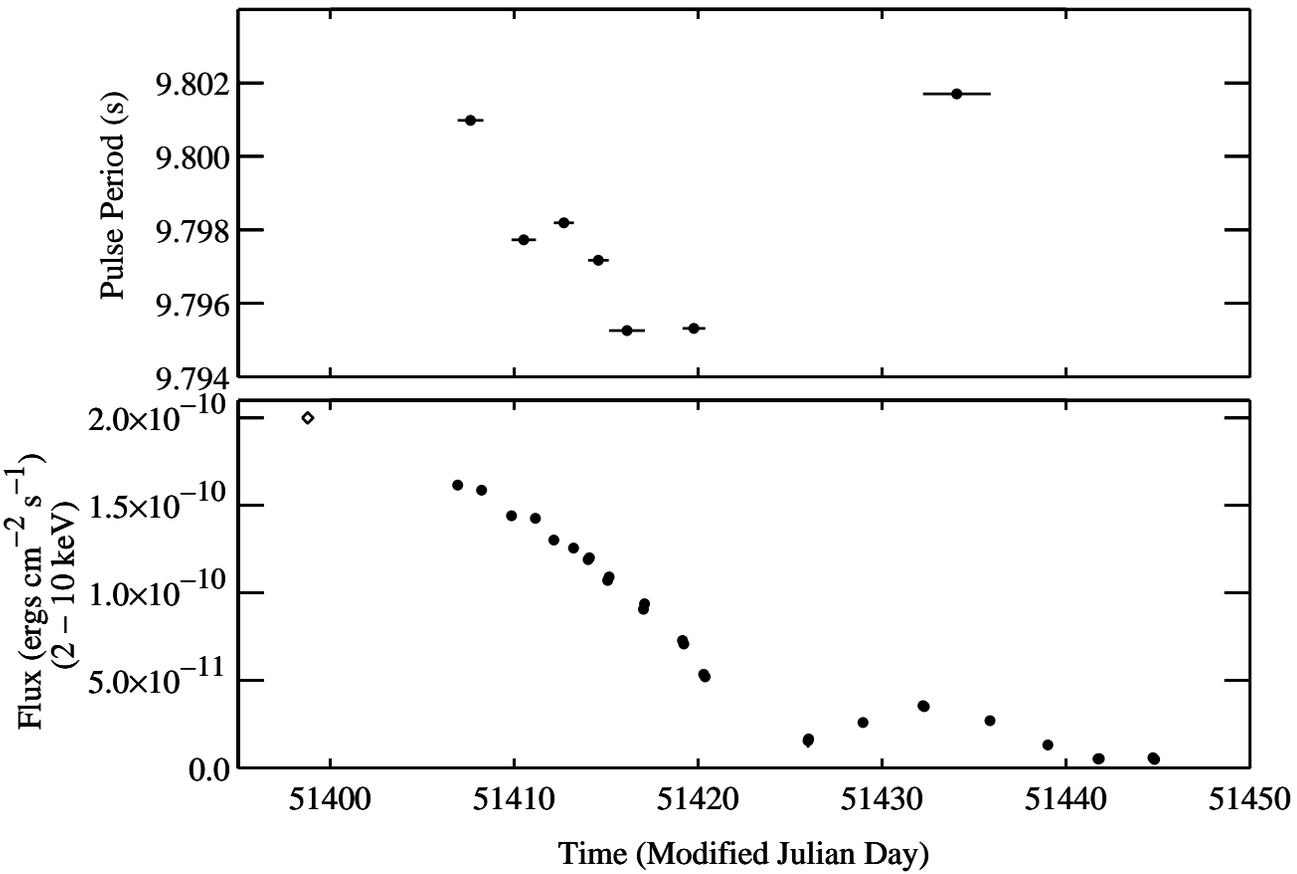}
\figcaption[f1.eps]{\RXTE\ PCA measurements
of the flux (lower panel) and pulse period (upper panel) of
\src. Pulse period measurements were not possible for
observations with low count rates or short durations. 
Flux and pulse period values are also
given in Tables \ref{table:spectrum} and \ref{table:pulse_period}
with the exception of the first flux measurement, marked
with a diamond, which is
taken from Marshall et al. (1999) and for which no error
is available.
The errors on the pulse period and flux
measurements are smaller than the symbol size.
\label{fig:pca_flux_period}
}
\end{figure}

The pulse timing results are given in Table \ref{table:pulse_period}
and plotted in Fig. \ref{fig:pca_flux_period}.
The period is found to change during the observations with an initial
decreasing trend in which the period changes are not monotonic
followed by an apparent period increase revealed by a single period measurement.The flux from \src\ is seen to initially decline by
a factor of more than 10
followed by a low amplitude
rebrightening at \sqig MJD 51,432.

\subsection{\RXTE\ ASM}

The \RXTE\ ASM (Levine et al. 1996) consists of three similar
Scanning Shadow Cameras
which perform sets of 90 second pointed
observations (``dwells'') so as to cover \sqig80\% of the sky every
\sqig90 minutes.  
Light curves are available in three energy bands: 1.5 to 3 keV, 3
to 5 keV, and 5 to 12 keV.
Source intensities are quoted as the count rates expected if the
source was in the center of the field of view of Scanning Shadow
Camera 1 in March 1996.  With this convention, the Crab Nebula has an
intensity of 75.5 \cts\ over the 1.5 - 12 keV energy range and
intensities of 26.8 (1.5 - 3 keV), 23.3 (3 - 5 keV), and 25.4 (5 - 12 keV)
\cts\ in each individual band.
Observations
of blank field regions away from the Galactic center suggest that
background subtraction may produce a systematic uncertainty of about 0.1
\cts\ (Remillard \& Levine 1997).
The ASM light curves used in our analysis 
span  the period from MJD 50,088 to 54,573
(1996 January 6 to 2008 April 17).

\begin{figure}
\epsscale{0.9}
\plotone{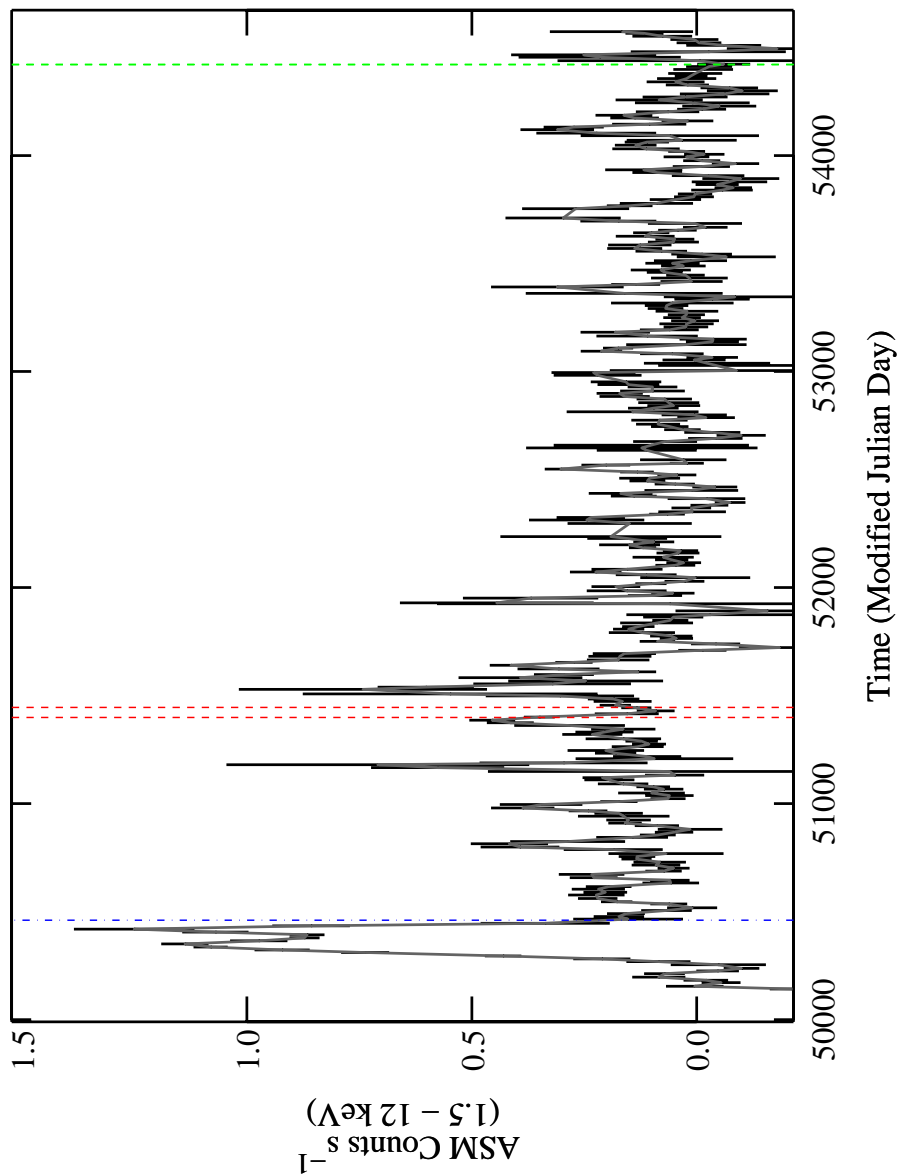}
\figcaption[f2.eps]{The \RXTE\ ASM light curve
of \src. The light curve has two week time bins which were then
smoothed using the algorithm described in
Section 2.2.
The vertical dot-dashed blue line indicates MJD 50,460,
before which the source may have experienced a ``Type II'' outburst.
The red vertical dashed lines at
MJD 51,398.78 and 51,444.82
indicate the interval
during which the PCA observations were made.
The green vertical dashed lines at MJD 54420.63 and 54421.98
indicate the time during which Swift observations failed to
detect the source (Romano et al. 2007).
\label{fig:asm_lc}
}
\end{figure}

The ASM light curve of \src\ is plotted in Fig. \ref{fig:asm_lc}.
For this light curve the dwell data were averaged
into 2 week time bins. That light curve was then
smoothed using an algorithm which replaces each data
point with a fraction of its original value plus fractions
of
the two immediately neighboring points. The factors used
for the relative contributions were 50\% of the original
value weighted by the inverse square of the error on that
value, plus 25\% of the neighboring points also weighted
by the inverse squares of their respective errors. 
The light curve
shows that an extended outburst occurred
between approximately MJD 50,250 to 50,460. 
The peak flux reached during this interval
was approximately 1 ASM \cts\ and we note
that during the outburst the flux never appears
to decrease to a low level. Excluding this period,
the mean flux is much lower at approximately 0.1 \cts.
If this source is a Be
star system, as suggested by Romano et al. (2007), then such
a long duration event is likely to be a ``Type II'' outburst during
which periodic modulation is not expected to be present
(e.g. Stella et al. 1986). 

\begin{figure}
\epsscale{0.9}
\plotone{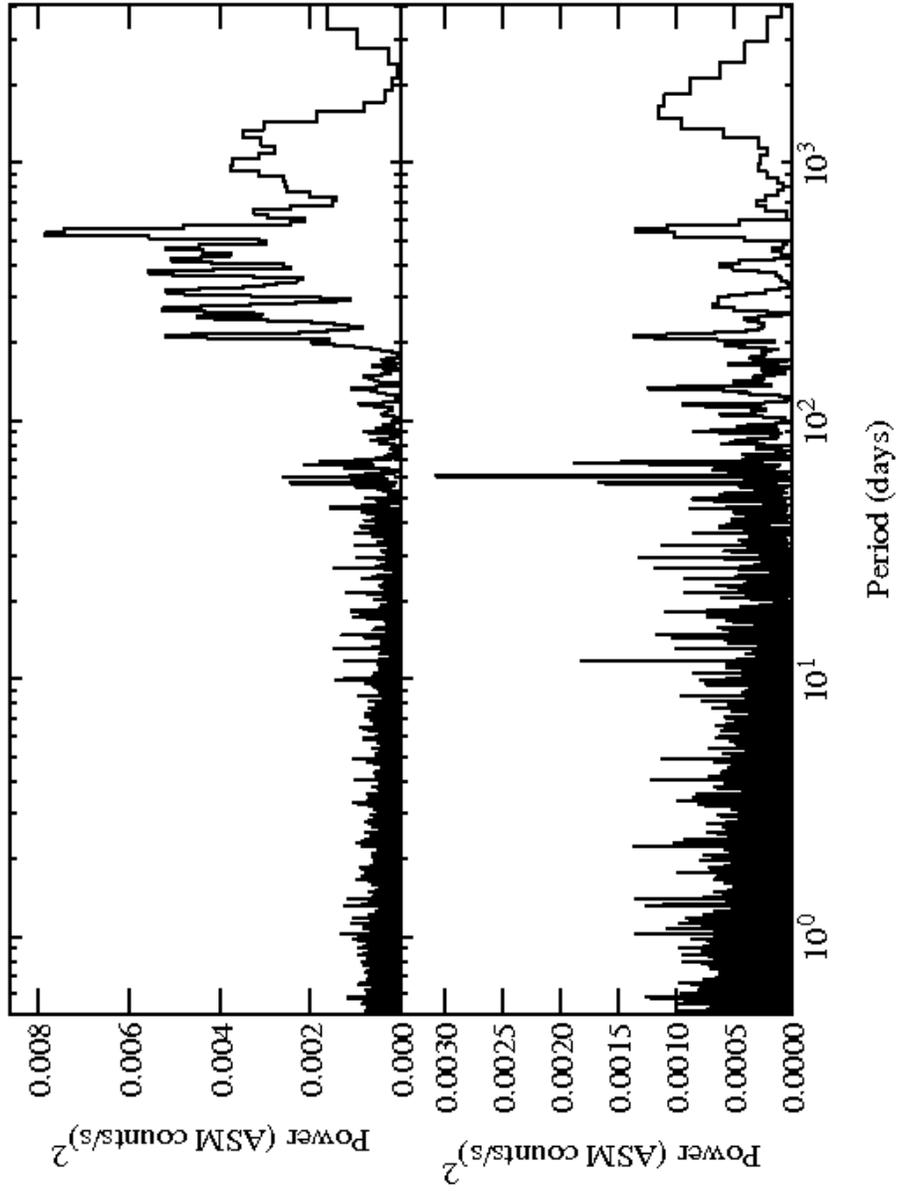}
\figcaption[f3.eps]{Power spectra of the \RXTE\ ASM
light curve of \src. The upper panel shows the power
spectrum of the entire light curve. The lower panel
shows the power spectrum using only observations
made after MJD 50,460.
\label{fig:1859_power}
}
\end{figure}

We computed power spectra of the light curve
to search for periodic modulation and
in all cases the calculation of power spectra employed the
``semi-weighting'' scheme discussed in Corbet et al. 
(2007a, b).
The power
spectrum of the entire light curve (Fig. \ref{fig:1859_power})
is dominated by low frequency noise
due to the extended outburst. However, the power spectrum of just the
ASM light curve obtained since MJD 50,460, after
the end of the outburst, shows a peak near a period
of 60.6 days (Fig.  \ref{fig:1859_power}).
The nominal false alarm probability (FAP) of obtaining
a peak with this strength or greater in the total frequency
range considered is
approximately 1\%. However, for the FAP to be completely
valid the number of independent trials, i.e. the
frequency range considered, must be decided before
a period is searched for (e.g. Scargle 1982).
The nominal significance
calculation also does not take into
account the difficult to quantify decision to exclude part of the data
and that the peak was first noted while searching
through ASM light curves of many sources and not just \src.
Therefore, the significance of the peak near 60.6 days is $<$99\%
and an exact figure cannot be given.
We also calculated power spectra of the light curves in
the 3 ASM energy bands. None of these power spectra showed
a modulation near 60.6 days that was more significant than
for the full energy range.
In order to further quantify the possible periodic modulation 
we fitted a sine wave to the light curve since MJD 50,460.
This gave a period of 60.65 $\pm$ 0.08 days with epoch
of maximum flux occurring at
MJD 52,549.1 $\pm$ 1.5. The fitted full amplitude of the modulation is
0.12 $\pm$ 0.02 \cts, equivalent to approximately 1.6 mCrab,
and the mean count rate is 0.092 $\pm$ 0.006 cts/s,
equivalent to approximately 1.2 mCrab.
The ASM light curve folded on a period of
60.65 days is shown in Fig. \ref{fig:1859_fold}.

\subsection{\BeppoSAX\ WFC}

The \BeppoSAX\ WFC instrument and its observing program are described by
Jager et al. (1997) and
Verrecchia et al. (2007) respectively.
The WFCs were two identical coded mask instruments
operating in the range 1.8 - 28 keV. The FOV was 40\degrees\
by 40\degrees\ with a FWHM angular resolution of 5'.
The source location accuracy for bright sources in crowded
fields was 0.7' at
99\% confidence level, and larger for sources detected
at lower significance levels.
The on-axis
sensitivity was between 2 to 10 mCrab for a typical \BeppoSAX\
WFC observation of 3$\times$10$^4$ s, and depended on the intensities of
other sources in the same FOV.
The WFCs pointed in opposite directions from each
other and at 90\degrees\ from the Narrow Field Instruments
also on board \BeppoSAX.
The WFCs operated between 1996 April 30
(MJD 50,203) and 2002 April 30 (MJD 52,394)
and covered the entire sky multiple times.

During 1996 to 2002 \src\ was in a BeppoSAX WFC FOV 115 times
for a total estimated net exposure time of 1.5 Msec. 
We searched for the combination of images that
produced the most significant detection of
\src. It was found that data taken
between MJD 50,340.65 and 50,395.91, i.e. during the period
when the ASM light curve indicates that \src\ was in an extended outburst,
gave the best results, and 
the signal-to-noise ratio is
16.7 for this interval. 
The count rate in the WFC was found to be too low
to allow useful spectroscopy to be obtained.
The best-fit source coordinates are:
R.A. = 18$^h$ 59$^m$ 2.4$^s$, decl. = +8\degrees\ 13' 57" (equinox 2000.0)
with
a 99\% confidence error radius of 1.6\arcmin. For a Gaussian error
distribution this translates to a 1.0\arcmin\ radius at a 90\%
confidence level. The equivalent Galactic coordinates are $l$ = 41.12\degrees, 
$b$ = 2.07\degrees.
This WFC error region has a four times smaller area than
the previously determined PCA position.
Outside this time interval we found no significant detection
of \src\ in individual observations or combinations of observations.
However, the WFC observations containing \src\ were
very non-uniform and there was more coverage during the earlier
part of the SAX mission, when the source was coincidentally
experiencing the extended outburst.

\begin{figure}
\epsscale{0.9}
\plottwo{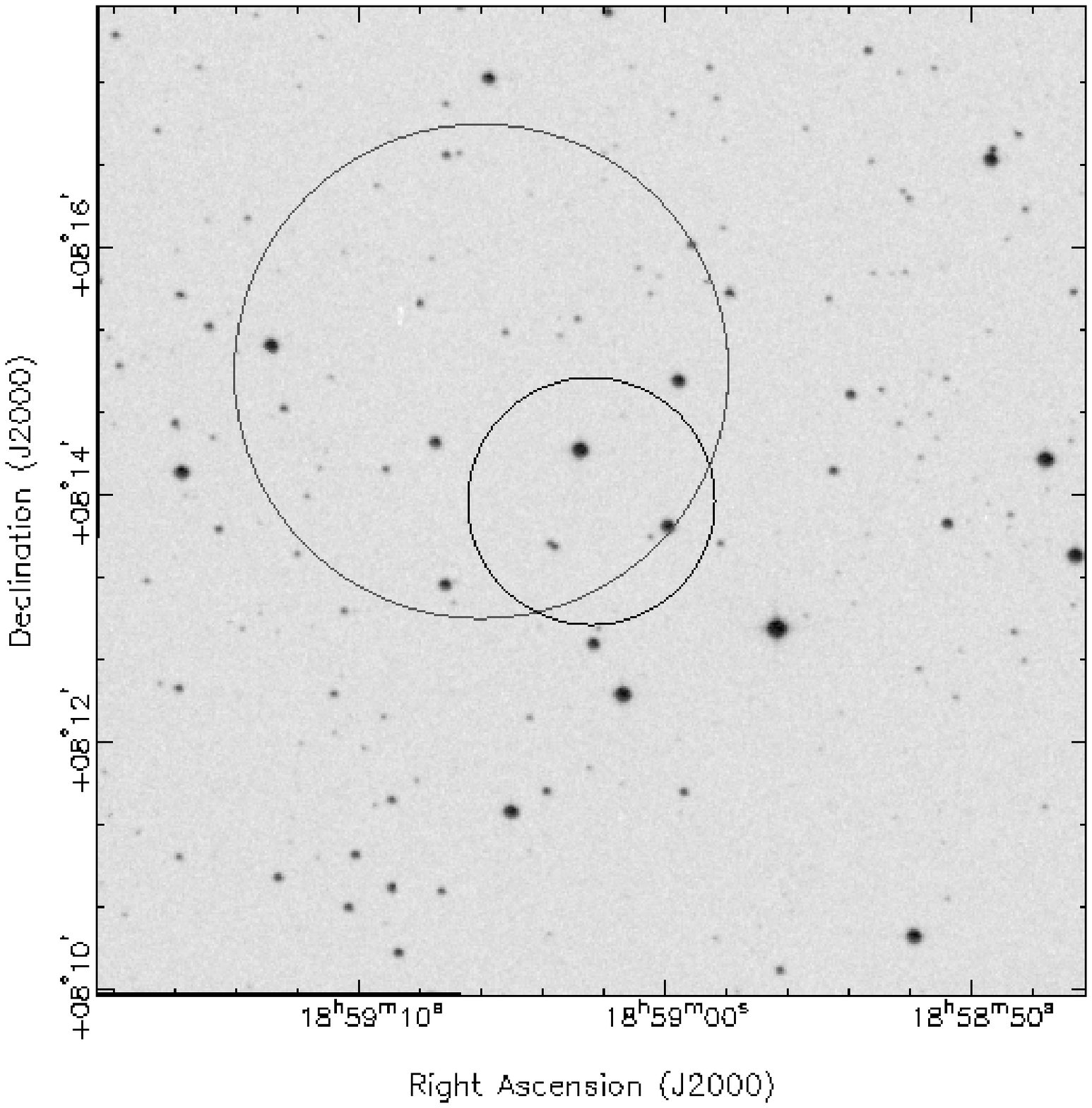}{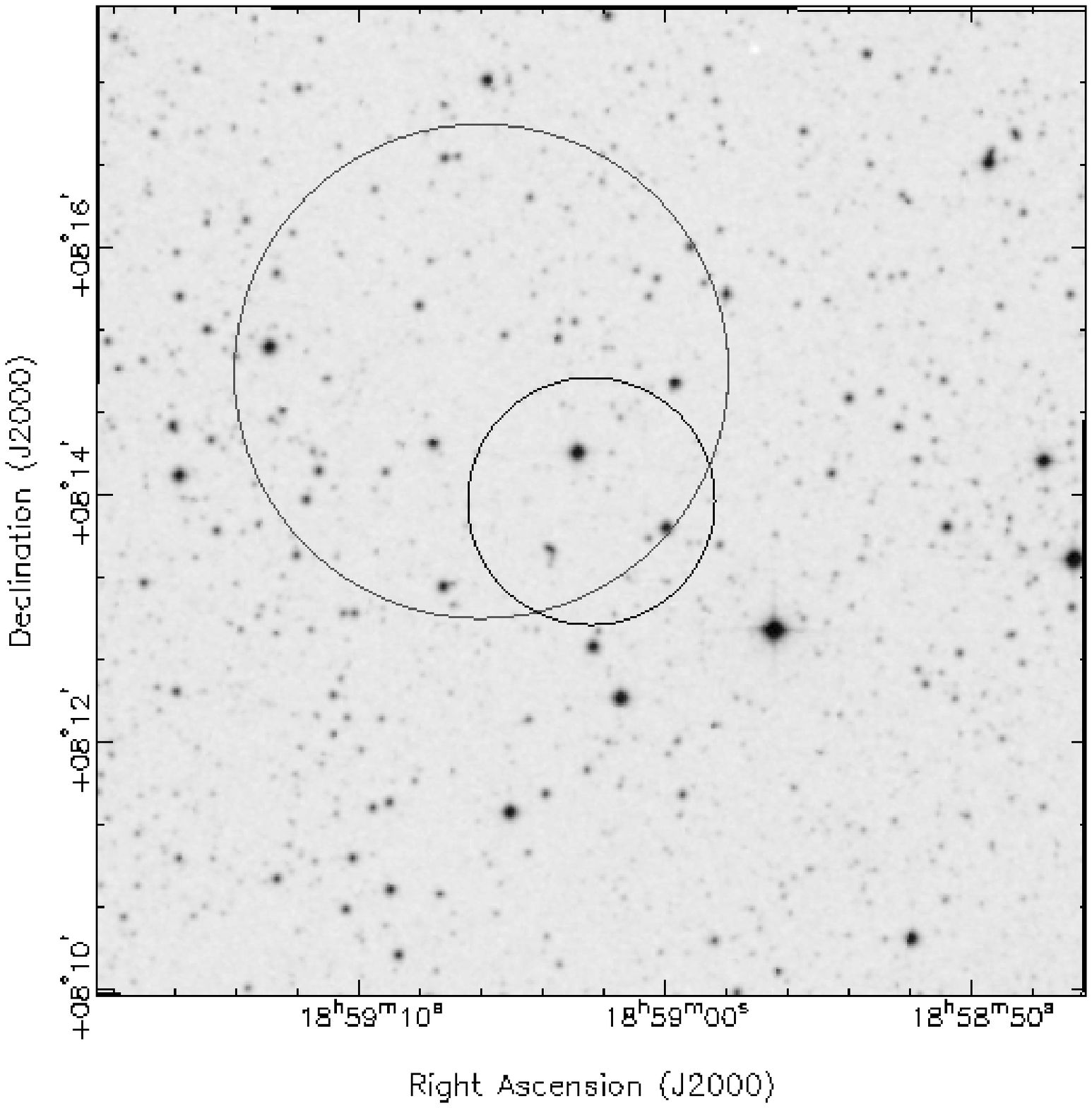}
\figcaption[f6.eps]{Error circles for the position
of \src\ on Digitized Sky Survey
images covering the DSS blue (left) and red (right)
bands.
North is up and East to the left.
The larger error circle to the North East
is the \RXTE\ PCA error 
region from Marshall et al. (1999) and the smaller circle
is from the \BeppoSAX\ WFC observations presented here.
\label{fig:dss}
}
\end{figure}

We plot both the PCA and WFC error regions on red
and blue Digitized
Sky Survey images in Fig.
\ref{fig:dss}. The DSS images are reported to be
complete down to $V$ = 21 (Monet et al. 2003).
Several stars are visible in the overlap of the
error boxes and the 
brightest object is 
USNO-B1.0 0982-0467446 ($B$ = 14.7)/2MASS
18590277+0814220 ($I$ = 11.7).
However, despite the reduction in the size of the
error box, without additional information on the
objects in the error box we cannot yet determine the optical
counterpart.

\section{Discussion}

A
60.65 day period would be consistent with the orbital period expected for a
Be star system containing a 9.8 s X-ray pulsar based on
the correlation between orbital and pulse periods for
this type of system (Corbet 1986).
However, the 60.65 day period is of modest 
statistical significance in
the ASM power spectrum. We therefore investigate whether
the PCA pulse timing
and flux measurements can help determine whether this
is indeed the orbital period of the system.

We first consider whether the changes in the pulse period
could be caused by accretion torques or orbital Doppler
shifts.
Joss \& Rappaport (1984) calculate that 
the spin-up rate can be approximately expressed
by: {$\dot{P}/P$} \sqig -3$\times$10$^{-5}$(P/1s)
(L$_X$/10$^{37}$erg s$^{-1}$)$^{6/7}$.
If the initial period decrease seen in \src\ was caused by
accretion torques, it thus requires a luminosity of about
10$^{39}$ \ergss. However, 
the brightest flux seen with the PCA only
corresponds to about 10$^{36}$ \ergss\ at 10 kpc.
Period change primarily caused by accretion torques thus appears to
be unlikely.

The mass function of a binary is given by
$f(M) = P_{orb}K^3(1 - e)^{3/2}/2\pi G.$
This corresponds to the minimum possible mass of the primary star.
We can estimate the mass function of \src\
for the assumptions that
the orbital period is 60.65 days and that the observed pulse period
changes are due to orbital motion.
Assuming a circular orbit for simplicity, the observed
maximum and minimum pulse periods imply a velocity
semi-amplitude of $\gtrsim$ 100 \kms\ which implies a mass function
of $\gtrsim$ 6 M\sun. The orbit is not fully sampled which
suggests the velocity amplitude should be larger than implied
by the limited pulse period measurements, and
hence that the mass function would be larger. Conversely,
if the orbit is significantly eccentric, as is common for
some Be star systems, then the mass function could be smaller.
Nevertheless, it appears that it is not unrealistic to
account for the observed pulse period changes by
orbital Doppler motion.
It is thus probable that orbital Doppler shifts were more important
than accretion torques in producing the observed pulse period changes.
We note, however, that the period changes are not completely
smooth. This suggests that accretion torques may be more
important than the simple estimate above implies.

We investigated whether it was possible to determine orbital
parameters by fitting the pulse period measurements.
Since there are only 7 period measurements, and the period increase
is only indicated by a single measurement, this severely limits
our ability to determine the orbital parameters.
Keeping the orbital period fixed at the value measured with
the ASM, we fitted a circular orbit. This resulted in an orbit
with a very large velocity
amplitude with a correspondingly unrealistically large mass function.
We conclude that more extensive pulse period measurements are
required before a reliable orbital solution can be determined.
In particular, period measurements should ideally span at
least one complete 60.65 day cycle.

\begin{figure}
\epsscale{0.9}
\plotone{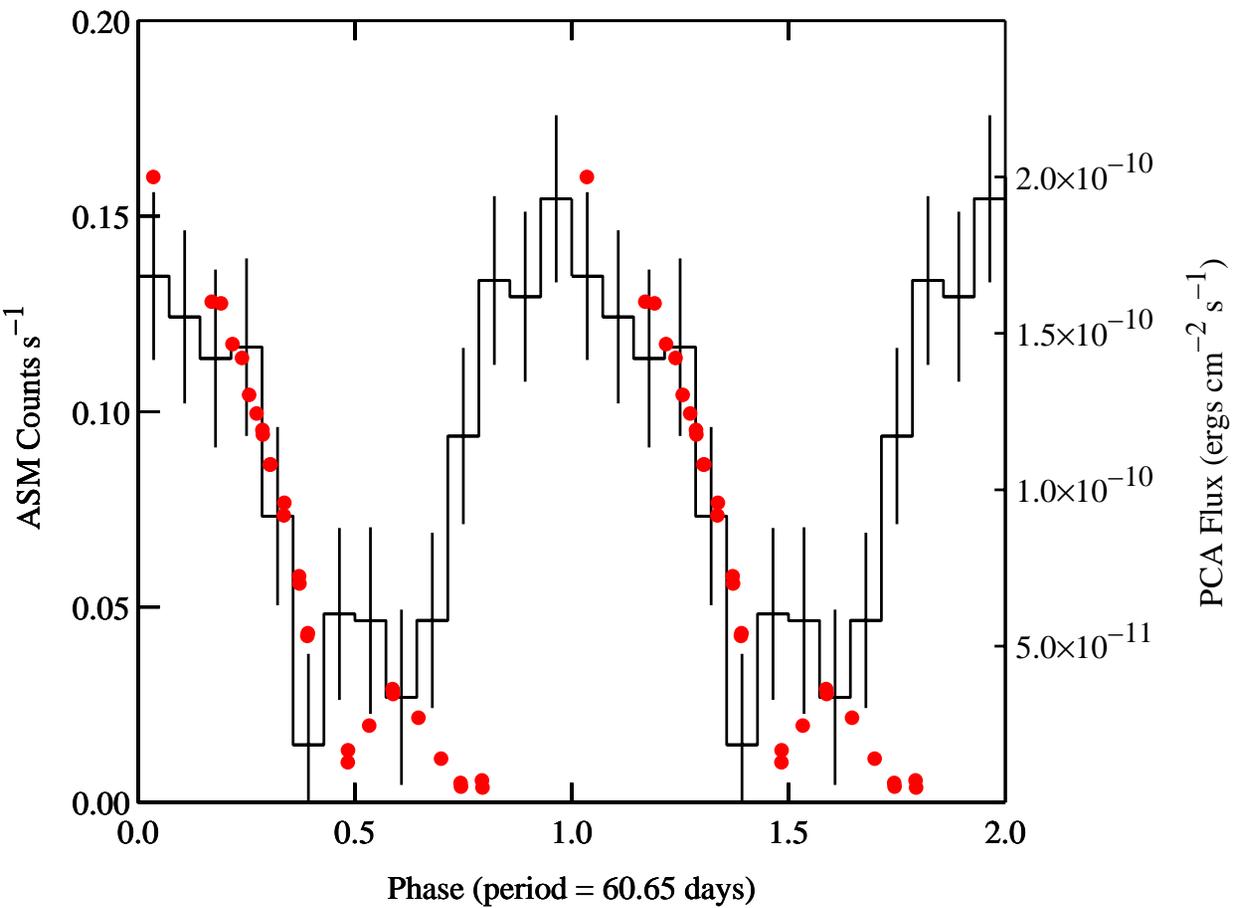}
\figcaption[f4.eps]{\RXTE\ ASM
light curve of \src\ since 
MJD 50,460 folded on the proposed
orbital period (histogram with error bars). 
The dots show the PCA fluxes scaled to the ASM light
curve to facilitate a comparison of the
morphology of the folded ASM light curve with the
PCA light curve. Note that the fluxes measured with the PCA
are much higher than the mean ASM flux 
(0.1 ASM \cts\ $\simeq$ 3.1$\times$10$^{-11}$ \ergss)
because
the ASM folded light curve also includes low flux states..
Phase zero corresponds to MJD 52,549.1.
\label{fig:1859_fold}
}
\end{figure}

\begin{figure}
\epsscale{0.9}
\plotone{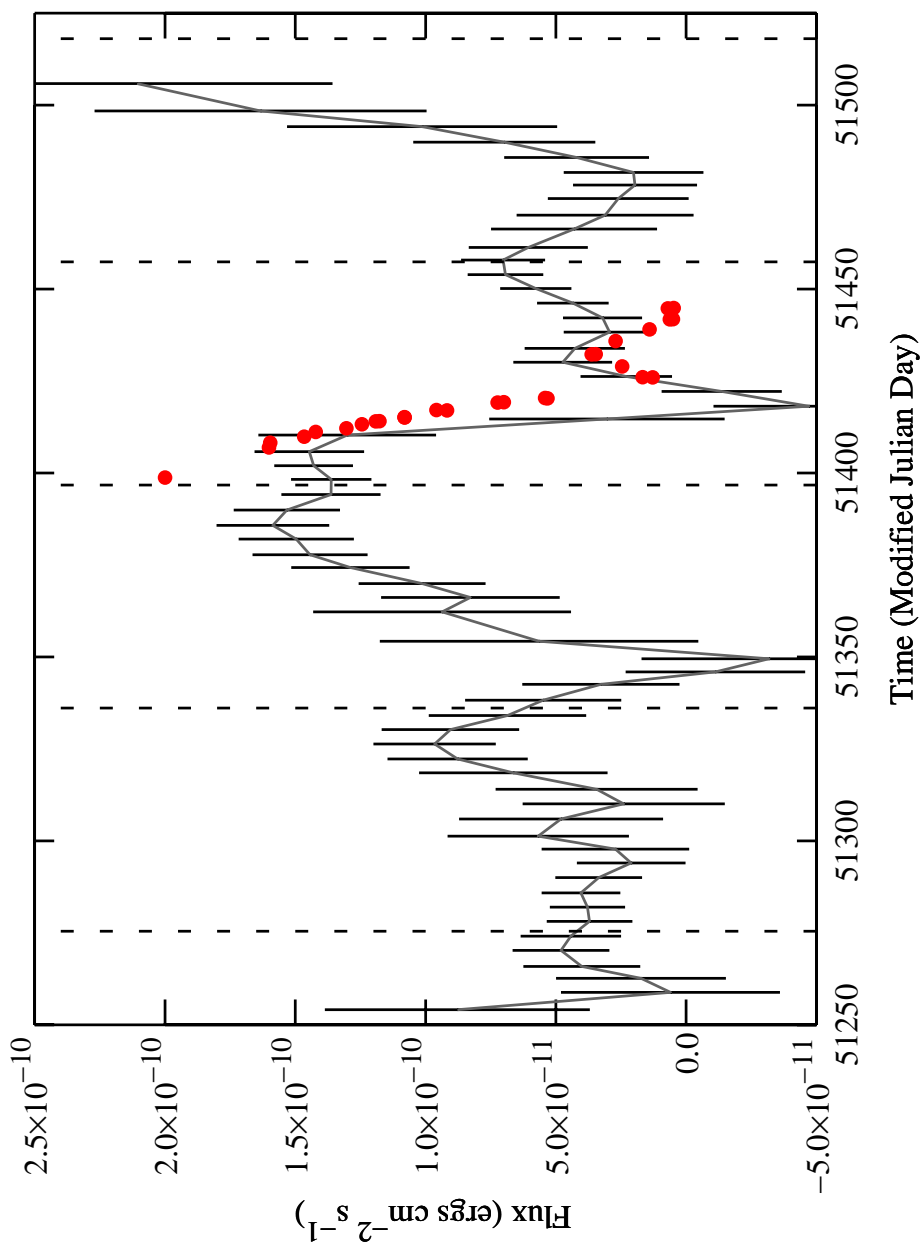}
\figcaption[f5.eps]{
ASM light curve of \src\ (lines with error bars)
around the time when the PCA observations were made.
The time resolution is 5 days and the light curve has
been smoothed using the algorithm described in
Section 2.2.
The red dots show the PCA fluxes.
The vertical dashed lines indicate the times
of predicted maximum flux based on the possible 60.65 day
period.
\label{fig:1859_lc_detail}
}
\end{figure}

We next consider whether the PCA flux modulation is
consistent with the proposed period.
In Fig. \ref{fig:1859_fold} we plot the PCA fluxes
on the folded ASM flux with the PCA peak flux arbitrarily
normalized to facilitate comparison of the
shape of the modulation. 
It is seen that the peak PCA flux and
decline from this have similar behavior to the folded 
ASM flux. However, the last few PCA observations
fall below the average ASM flux.
In Fig. \ref{fig:1859_lc_detail} we overplot the PCA
fluxes and the ASM fluxes obtained during the same
time.
Due to the lower sensitivity of the ASM compared to
the PCA a 5 day time resolution was used for the
ASM light curve and this was then smoothed using
the same algorithm described in Section 2.2.
We convert the ASM count rate to flux using the spectrum
derived in Section 2.1 which gives a conversion
factor of 0.1 ASM \cts\ $\simeq$ 3.1$\times$10$^{-11}$ \ergscms.
There is reasonable agreement of the flux level
seen in the two instruments with the exception that
the flux measured from the slew observations appears brighter
than that measured with the ASM. Note, however, that the
PCA flux is derived from a very brief observation whereas
the ASM light curve uses smoothed 5 day bins.
The ASM data suggest
that a modest outburst at the time of the next
predicted maximum occurred shortly after
the PCA observations ceased.
It thus appears that the light curve obtained
with the PCA is consistent with the proposed
60.65 day modulation.
If the primary of the system is a Be star then
modulation of the light curve on the orbital period
would
depend on the structure of the envelope around
the Be star, i.e. its density and velocity
and the orbit of the neutron star. Variability
of the properties of the Be star envelope, as would
be expected to occur, would
lead to changes in the orbital modulation of the light
curve. If the orbital parameters of \src\ can be confirmed,
and the optical counterpart identified, then it may be possible
to derive parameters of the Be envelope from the modulation
of the light curve and emission at H$\alpha$ which would
come from the envelope and hence also provide information
on the physical conditions of the envelope.

We note that the {\it Swift} 
non-detection of \src\ (Romano et al. 2007) occurred at a phase of
0.85. The folded ASM light curve 
(Fig. \ref{fig:1859_fold})
shows that \src\ has generally
been bright at that phase, suggesting that the source may have been in
an extended inactive state. The occurrence of extended inactive states
where the outbursts near periastron passage cease is common
for Be/neutron star binaries (e.g. Charles \& Coe 2006).

\section{Conclusion}

The spectral, flux variability, and timing 
properties of \src\ appear to be consistent with
a Be star X-ray binary classification.
\src\ appears to have exhibited the three types of
emission state commonly seen in this type of source:
(i) a bright Type II outburst with no periodic modulation;
(ii) a lower luminosity state during which the PCA observations
were made and during which periodic modulation of the
flux may have occurred; and (iii) a quiescent state
during which no emission was detectable.
Additional observations to
further reduce the size of the error box would be valuable,
and these should preferably be carried out
at the peak of the candidate 60.65 day period.
If the source enters another extended bright phase
then additional pulse timing measurements should be able
to fully determine
the orbital parameters of this system.

\acknowledgements
We thank Cees Bassa for making the DSS finding charts.

\begin{deluxetable}{lccccccccccc}
\rotate
\tablewidth{8.2in}
\tabletypesize{\scriptsize}
\tablecaption{\RXTE\ PCA Observations of \src}
\tablehead{

\colhead{ID} & \colhead{Start} & \colhead{End} & \colhead{Length} & \colhead{2 - 10 keV Flux} & 
\colhead{Photon}  & \colhead{$N_{\rm H}$} & $E_{\rm cut}$ & $E_{\rm fold}$ 
& \colhead{Count Rate} 
 & \colhead{$\chi^2_\nu$} & \colhead{Phase}\\

\colhead{} & \colhead{Time} & \colhead{Time} & \colhead{(s)}  &\colhead{(\ergscms)$\times$10$^{-11}$} &
\colhead{Index}  & \colhead{(10$^{22}$cm$^{-2})$}  & (keV) & (keV) & \colhead{} 
 & \colhead{} &  \colhead{Range}\\

}
\startdata
1  &   6.93272 & 6.98342 & 1184   & 16.15 +0.10, -0.09  & 0.93 $\pm$  0.21 &  2.98 $\pm$  1.58 &6.75 $\pm$   0.36  & 13.30 $\pm$   0.88 & 68.02 $\pm$ 0.45 & 0.83 & 0.1695 - 0.1704 \\
2  &   8.23485 & 8.33245 & 4976   & 15.86 +0.04, -0.05  & 0.75 $\pm$  0.11 &  1.53 $\pm$  0.76 &6.39 $\pm$   0.20  & 13.54 $\pm$   0.52 & 64.50 $\pm$ 0.23 & 1.06 & 0.1910 - 0.1926 \\
3  &   9.85825 & 9.87297 & 1072   & 14.39 +0.12, -0.11  & 0.76 $\pm$  0.31 &  0.95 $\pm$  2.32 &7.15 $\pm$   0.50  & 15.07 $\pm$   1.37 & 41.83 $\pm$ 0.39 & 0.47 & 0.2178 - 0.2180 \\
4  &   11.16430 & 11.18726 & 1952 & 14.25 +0.07, -0.06  & 0.91 $\pm$  0.19 &  2.54 $\pm$  1.42 &6.63 $\pm$   0.30  & 13.69 $\pm$   0.77 & 73.93 $\pm$ 0.41 & 0.91 & 0.2393 - 0.2397 \\
5  &   12.16282 & 12.18671 & 1984 & 13.01 +0.07, -0.06  & 0.69 $\pm$  0.16 &  1.54 $\pm$  0.99 &6.05 $\pm$   0.30  & 12.12 $\pm$   0.78 & 70.65 $\pm$ 0.40 & 0.77 & 0.2558 - 0.2562 \\
6  &   13.22880 & 13.25615 & 2144 & 12.55 +0.06, -0.07  & 0.95 $\pm$  0.22 &  2.99 $\pm$  1.52 &6.28 $\pm$   0.36  & 13.37 $\pm$   0.96 & 50.73 $\pm$ 0.33 & 0.44 & 0.2734 - 0.2738 \\
7  &   14.03024 & 14.04171 &  976 & 11.89 +0.11, -0.08  & 0.75 $\pm$  0.31 &  1.72 $\pm$  2.04 &6.30 $\pm$   0.50  & 12.26 $\pm$   1.23 & 47.43 $\pm$ 0.47 & 1.21 & 0.2866 - 0.2868 \\
8  &   14.09300 & 14.11300 & 1568 & 12.00 +0.09, -0.07  & 0.99 $\pm$  0.27 &  3.85 $\pm$  1.91 &6.29 $\pm$   0.39  & 12.11 $\pm$   0.95 & 47.73 $\pm$ 0.38 & 0.80 & 0.2876 - 0.2879 \\
9  &   15.09189 & 15.14485 & 1584 & 10.70 +0.12, -0.08  & 0.97 $\pm$  0.33 &  2.09 $\pm$  2.40 &6.61 $\pm$   0.60  & 15.92 $\pm$   1.87 & 29.44 $\pm$ 0.31 & 0.83 & 0.3041 - 0.3049 \\           
10  &  15.17253 & 15.18319 &  912 & 10.90 +0.09, -0.10  & 0.98 $\pm$  0.30 &  3.04 $\pm$  2.00 &6.14 $\pm$   0.51  & 13.36 $\pm$   1.41 & 71.33 $\pm$ 0.64 & 0.62 & 0.3054 - 0.3056 \\
11  &  17.02948 & 17.03893 &  800 &  9.05 +0.13, -0.15  & 0.90 $\pm$  0.47 &  1.63 $\pm$  3.68 &7.00 $\pm$   0.74  & 13.34 $\pm$   1.89 & 24.39 $\pm$ 0.42 & 0.90 & 0.3360 - 0.3362 \\
12  &  17.08985 & 17.11022 & 1584 &  9.37 +0.11, -0.11  & 0.73 $\pm$  0.35 &  0.65 $\pm$  2.42 &6.59 $\pm$   0.57  & 13.61 $\pm$   1.54 & 25.22 $\pm$ 0.30 & 0.98 & 0.3370 - 0.3373 \\           
13  &  19.15636 & 19.17852 & 1888 &  7.27 +0.07, -0.08  & 0.99 $\pm$  0.31 &  2.93 $\pm$  2.10 &6.21 $\pm$   0.48  & 12.59 $\pm$   1.30 & 37.37 $\pm$ 0.38 & 1.08 & 0.3711 - 0.3715 \\
14  &  19.22750 & 19.25004 & 1920 &  7.07 +0.07, -0.07  & 1.02 $\pm$  0.33 &  3.27 $\pm$  2.35 &6.25 $\pm$   0.46  & 12.16 $\pm$   1.19 & 36.46 $\pm$ 0.37 & 0.92 & 0.3723 - 0.3726 \\
15  &  20.31178 & 20.32995 &  752 &  5.34 +0.14, -0.13  & 0.95 $\pm$  0.74 &  2.43 $\pm$  4.88 &6.27 $\pm$   0.92  & 10.73 $\pm$   2.08 & 20.27 $\pm$ 0.49 & 0.72 & 0.3901 - 0.3904 \\
16  &  20.38468 & 20.39592 &  832 &  5.20 +0.15, -0.16  & 1.00 $\pm$  0.81 &  0.89 $\pm$  5.49 &6.40 $\pm$   1.37  & 14.52 $\pm$   4.17 & 13.73 $\pm$ 0.38 & 0.85 & 0.3913 - 0.3915 \\           
17  &  25.97901 & 25.99036 &  128 &  1.54 +0.32, -0.39  & 1.17 $\pm$  7.20 & 20.97 $\pm$ 52.43 &6.45 $\pm$   2.48  &  3.53 $\pm$   2.58 &  5.15 $\pm$ 1.23 & 0.56 & 0.4836 - 0.4838 \\
18  &  26.01698 & 26.02911 & 1040 &  1.64 +0.13, -0.12  & 1.42 $\pm$  1.99 &  3.39 $\pm$ 13.01 &6.67 $\pm$   2.64  & 11.82 $\pm$   7.15 &  5.80 $\pm$ 0.42 & 0.48 & 0.4842 - 0.4844 \\
19  &  28.97416 & 29.02578 & 1280 &  2.58 +0.11, -0.11  & 1.81 $\pm$  0.95 &  7.94 $\pm$  7.20 &6.13 $\pm$   1.03  &  7.98 $\pm$   1.89 &  9.40 $\pm$ 0.38 & 0.50 & 0.5330 - 0.5338 \\
20  &  32.23217 & 32.24102 &  768 &  3.54 +0.15, -0.15  & 0.97 $\pm$  0.81 &  1.34 $\pm$  4.85 &5.76 $\pm$   2.57  & 17.94 $\pm$  10.26 & 16.20 $\pm$ 0.50 & 0.53 & 0.5867 - 0.5868 \\
21  &  32.30152 & 32.31536 & 1056 &  3.49 +0.13, -0.12  & 1.15 $\pm$  0.85 &  3.23 $\pm$  6.21 &6.73 $\pm$   2.09  & 20.41 $\pm$   8.23 & 17.16 $\pm$ 0.43 & 1.12 & 0.5878 - 0.5881 \\
22  &  35.87078 & 35.91116 & 1008 &  2.69 +0.12, -0.12  & 1.00 $\pm$  1.10 &  2.52 $\pm$  6.69 &6.14 $\pm$   1.43  & 10.49 $\pm$   3.29 &  9.37 $\pm$ 0.41 & 0.90 & 0.6467 - 0.6473 \\
23  &  39.01763 & 39.03967 & 1808 &  1.30 +0.09, -0.11  & 1.22 $\pm$  1.84 &  0.37 $\pm$ 11.24 &6.44 $\pm$   1.67  &  9.31 $\pm$   3.49 &  4.51 $\pm$ 0.30 & 1.01 & 0.6986 - 0.6989 \\
24  &  41.74263 & 41.74866 &  528 &  0.51 +0.09, -0.14  & 0.56 $\pm$ 10.04 &  0.43 $\pm$ 32.94 &6.19 $\pm$   3.05  &  3.96 $\pm$   4.42 &  2.51 $\pm$ 0.66 & 0.69 & 0.7435 - 0.7436 \\
25  &  41.81208 & 41.83269 & 1744 &  0.52 +0.09, -0.10  & 1.80 $\pm$  3.09 &  5.67 $\pm$ 22.59 &6.79 $\pm$   1.02  &  1.68 $\pm$   1.00 &  0.67 $\pm$ 0.30 & 0.90 & 0.7446 - 0.7450 \\
26  &  44.73542 & 44.74588 &  896 &  0.57 +0.09, -0.12  & 2.37 $\pm$  3.94 &  1.52 $\pm$ 20.96 &3.65 $\pm$  10.32  &  9.99 $\pm$  34.44 &  2.52 $\pm$ 0.43 & 0.73 & 0.7928 - 0.7930 \\
27  &  44.82212 & 44.83272 &  816 &  0.47 +0.09, -0.14  & 0.89 $\pm$  3.27 &  0.00 $\pm$ 52.25 &7.31 $\pm$   1.28  &  0.73 $\pm$   1.25 & -0.58 $\pm$ 0.43 & 1.00 & 0.7943 - 0.7944 \\
\enddata
\tablecomments{ 
Start and end times are in units of MJD - 51,400.
The flux is not corrected for absorption. Flux errors are
1$\sigma$ confidence intervals obtained from fitting the spectra with all parameters fixed
at the mean values from fitting all spectra, with the exception of the normalization and $E_{\rm fold}$.
The phase range is for a period of 60.65 days and phase zero corresponds to MJD 52,549.1.
}	
\label{table:spectrum}
\end{deluxetable}

\clearpage

\begin{deluxetable}{lccc}
\tablecaption{\RXTE\ PCA Pulse Period Measurements of \src}
\tablehead{
\colhead{Observation} & \colhead{Start Time} & 
\colhead{End Time} & \colhead{Period} \\
\colhead{Numbers} & \colhead{(MJD)} & 
\colhead{(MJD)} & \colhead{(s)} \\
}

\startdata
1, 2 &  51406.93272 & 51408.33245   & 9.80098 $\pm$ 0.00002 \\
3, 4 &  51409.85825 &  51411.18726            & 9.79773 $\pm$ 0.00001 \\
5, 6 &  51412.16282 &  51413.25615             & 9.79819 $\pm$ 0.00003 \\
7, 8, 9 &  51414.03024 &   51415.14485             & 9.79717 $\pm$ 0.00001 \\
10, 12 & 51415.17253 &   51417.11022         & 9.79526 $\pm$ 0.00001 \\
13, 16 & 51419.15636 &   51420.39592         & 9.79532 $\pm$ 0.00003 \\
20, 21, 22 & 51432.23217 &   51435.91116     & 9.80170 $\pm$ 0.00001 \\
\enddata
\label{table:pulse_period}
\end{deluxetable}

\clearpage

\end{document}